\documentclass{cfdsc}
\usepackage[T1]{fontenc}
\usepackage{setspace}
\usepackage{amssymb}
\usepackage{hyperref}

\usepackage{color}

\usepackage{caption}
\usepackage{subcaption}

\begin{document}

\title{Flow Regimes of Mesoscale Circulations Forced by Inhomogeneous Surface Heating}

\author{M. Alamgir Hossain \and Jahrul M Alam\thanks{Correspondence to: alamj@mun.ca}}

\institute{Department of Mathematics and Statistics, Memorial University of Newfoundland, Canada}

\email{mah782@mun.ca and alamj@mun.ca}

\maketitle

\begin{abstract}
Urbanization is one of the extreme process that increases uncertainty in future climate projections. Flow regimes of mesoscale circulations associated with surface heating due to urbanization have been investigated using a wavelet based computational fluid dynamics~(CFD) model. 
The results of our numerical model have been validated against that of a laboratory model, as well as reference numerical simulations. Characteristics of urban induced circulations have been studied for surface heat flux perturbation ($H_0$) between $28.93$~W~m$^{-2}$ and $925.92$~W~m$^{-2}$, and the results have been analyzed against available boundary layer measurements under similar physical conditions. Our primary study shows that urban/rural heat flux anomalies introduce strong oscillations in the convective boundary layer (CBL), and transfers a fraction of the turbulent kinetic energy vertically through internal waves. Such results complement previous investigators' hypothesis that temporal oscillations in urban-induced mesoscale circulations are primarily due to a downscale energy cascade. Although a further detailed study is necessary, the present numerical observations provide useful feedback for the impacts of urbanization on regional climate.
\end{abstract}

\begin{figure*}[t]
\noindent\includegraphics[width=\textwidth]{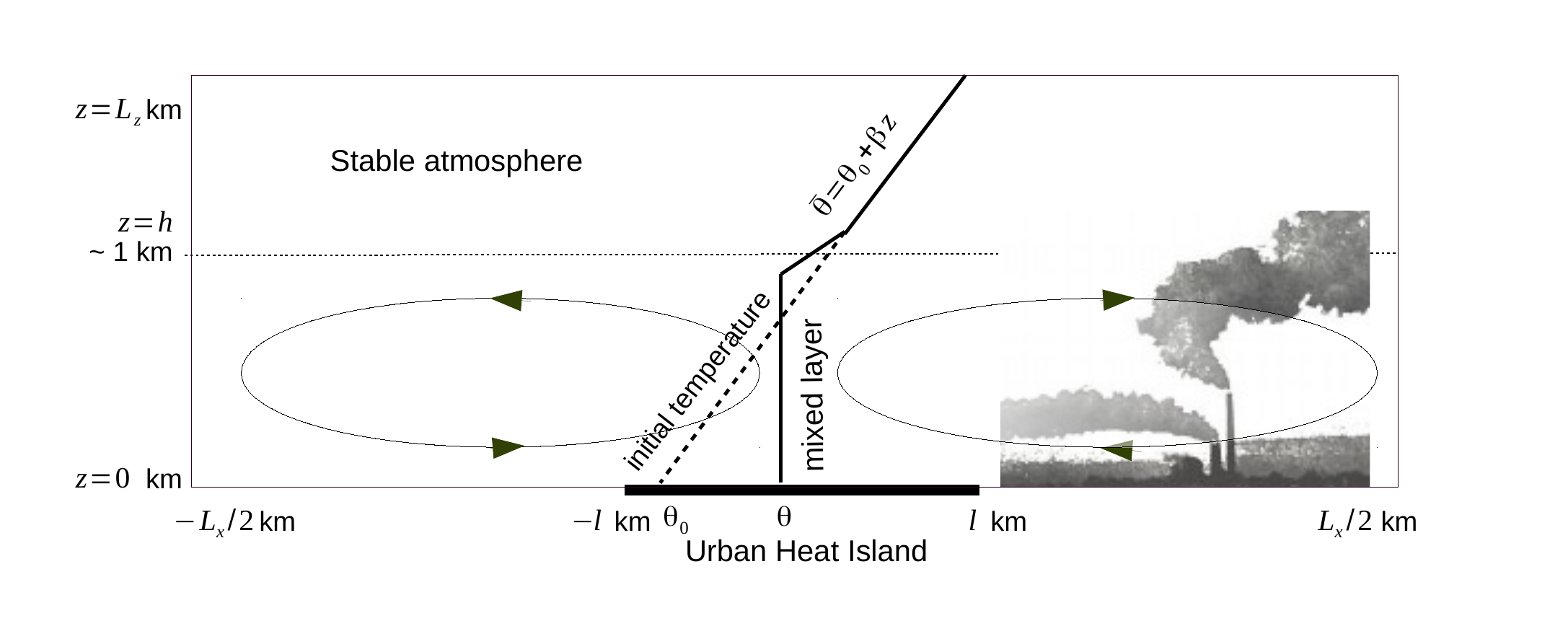}\\
\caption{Schematic representation of potential temperature profiles in convective boundary layer. The thick black horizontal line represents the urban heated area. (Photo by Ralph Turncote)\label{fig:uhisv}}
\end{figure*}

\section{Introduction}
Human modification of the earth's surface makes city areas significantly warmer than their surrounding rural areas during the daytime, when the sun heats the surface. Usually, cities have a mean temperature that is 8 to 10 degrees higher than the surrounding rural areas ({\em e.g.} \cite{tran2006assessment}). The air circulation due to this heat source is known as urban heat island (UHI) circulation. Surface heterogeneity due to urbanization usually triggers mesoscale circulations, which influence boundary layer meteorology and atmospheric turbulence \cite{collier2006impact}. For example, observational data analysis indicates that there is a strong urban induced influence on local storms, precipitation, convective rolls, lake-breeze, heatwave etc. ({\em e.g.} \cite{niyogi2011urban}). The increase of clear-air turbulence in the North Atlantic, USA, and European sectors by 40-90\% over the period of 1958-2001 clearly demands a detailed investigation of how urbanization enhances vertical transport and mixing through internal waves \cite{jaeger2007northern}. Because of its importance in the environmental problems, theoretical models ({\em e.g.} \cite{Nino2005, ueda1983effects}), laboratory experiments ({\em e.g.} \cite{Kimura1975}), observational studies ({\em e.g.} \cite{childs2005observations, moeng1984large}), and numerical simulations ({\em e.g.} \cite{baik2001dry, delage1970numerical,  Dubois2009, richiardone1989numerical, Zhang2014}), were considered. 

An interesting feature found from laboratory \cite{Kimura1975} and numerical studies \cite{Nino2005} of UHI flows is that the heat island circulation has two types of flow regimes, which depend on the intensity of inhomogeneous heating. A numerical study of heat island flows at Rayleigh number, $Ra \le 10^5$ is presented by Dubois and Touzani \cite{Dubois2009}. In their simulations, a bounded elongated domain was used, where they implemented a thermal sponge layer in the vicinity of the vertical boundaries.  Zhang {\em et al.} \cite{Zhang2014} used the large-eddy simulation mode of the WRF model to investigate mesoscale circulation that is induced by inhomogeneous surface heating. Turbulent heat fluxes in urban areas are investigated by \cite{Grimmond2002}. Stull \cite{stull1976internal} and Deardorff \cite{Deardorff1970} investigated penetrative turbulence in oceanic and atmospheric boundary layers. Such investigations may further be extended through numerical simulation. Recently, Hossain \cite{Alamgir2015} has studied the penetrative turbulence in convective boundary layers generated by a differential heating in a large horizontal surface. The impact of surface heating on mesoscale circulation and boundary layer meteorology is not fully understood and the topic is a active research area, where CFD techniques can adopt for investigating the characteristics of such flows.

This paper presents a Computational Fluid Dynamics model for the numerical simulation of mesoscale circulations triggered by surface heat flux, which also illustrates the coherent flow regime associated with penetrative turbulence. Our simulations indicate that a strong interaction exists between urban heating and clear-air turbulent transport through the mechanism of internal waves. However, this article has not fully characterized the regime of internal wave formation.

\section{Governing Equations}
Consider the ground surface is flat, the influence of the terrain is ignored, the flow is uniform along the spanwise direction, and the Coriolis force is negligible. Under these assumptions, the governing equations of the compressible dry atmosphere are 
\begin{equation}
 \frac{\partial \pi}{\partial t} + u_j \frac{\partial \pi}{\partial x_j} = - \pi \frac{\partial u_i}{\partial x_i}, 
\label{eq:ch2_28}
\end{equation}
\begin{equation}
\frac{\partial u_i}{\partial t} +  u_j \frac{\partial u_i}{\partial x_j} = - \theta_0 \frac{\partial \pi}{\partial x_i} + \frac{\partial \tau_{ij}}{\partial x_j} + \frac{\theta}{\theta_0} g \delta_{i3}, 
\label{eq:ch2_29}
\end{equation}
\begin{equation} 
\frac{\partial \theta}{\partial t} + u_j \frac{\partial \theta}{\partial x_j}+ w \beta = \frac{\partial\tau_{\theta j}}{\partial x_j},   
\label{eq:ch2_30}
\end{equation}
where $\beta = \partial \bar \theta/\partial z$, $\delta_{ij}$ is the Kronecker delta and $\tau_{ij}$ is the turbulent momentum flux and $\tau_{\theta j}$ is the turbulent heat flux. The nondimensional pressure ($\pi$) is defined as 
\[\pi = \left( \frac{p}{p_0}\right)^{R/C_p} \]
where $p_0 = 1000$~mb and $c_p = 1004$ J kg$^{-1}$ K$^{-1}$ is the specific heat of dry air at constant pressure. Subgrid scale stress terms in eq.~(\ref{eq:ch2_29}) and (\ref{eq:ch2_30}) can be parametrized as 

\[ \frac{\partial \tau_{ij}}{\partial x_j} = \frac{\partial}{\partial x} \left( K_M \frac{\partial u_i}{\partial x}\right) + \frac{\partial}{\partial z} \left(K_m\frac{\partial u_i}{\partial z} \right),\]

\[\frac{\partial \tau_{\theta j}}{\partial x_j} = \frac{\partial}{\partial x} \left( K_M \frac{\partial \theta}{\partial x}\right) + \frac{\partial}{\partial z} \left(K_{H}\frac{\partial \theta}{\partial z} \right), \]
where $K_M$ and $K_m$ are horizontal and vertical eddy viscosity, respectively, for momentum, and $K_{H}$ is the eddy diffusivity of heat. The large eddy simulation (LES) model described in Deardorff \cite{Deardorff1970} is implemented to estimate the coefficients $K_M$, $K_m$ and $K_H$. If the scale ($\Delta$) of resolved large eddies is within the inertial range, then
\[  K = C_s^{4/3}~ \bar \epsilon^{1/3}~ \Delta^{4/3},\] 
where subscripts are dropped for simplicity \cite{Deardorff1970}, $C_s$ is a dimensionless constant which is often called Smagorinsky constant, $\bar \epsilon$ is the rate of dissipation of turbulent energy within a local grid volume, and  the filtered scale is defined by  $\Delta = (\Delta x~\Delta y~\Delta z)^{1/3}$, where $\Delta x,~\Delta y~ \textrm{and}~\Delta z$ are the length of $x$, $y$ and $z$ direction, respectively, of the filtered cell.

\begin{figure*}
  \centering             
 \begin{subfigure}[b]{0.31\textwidth}
   \includegraphics[trim=1cm 0cm 1.5cm 0cm, width=\textwidth]{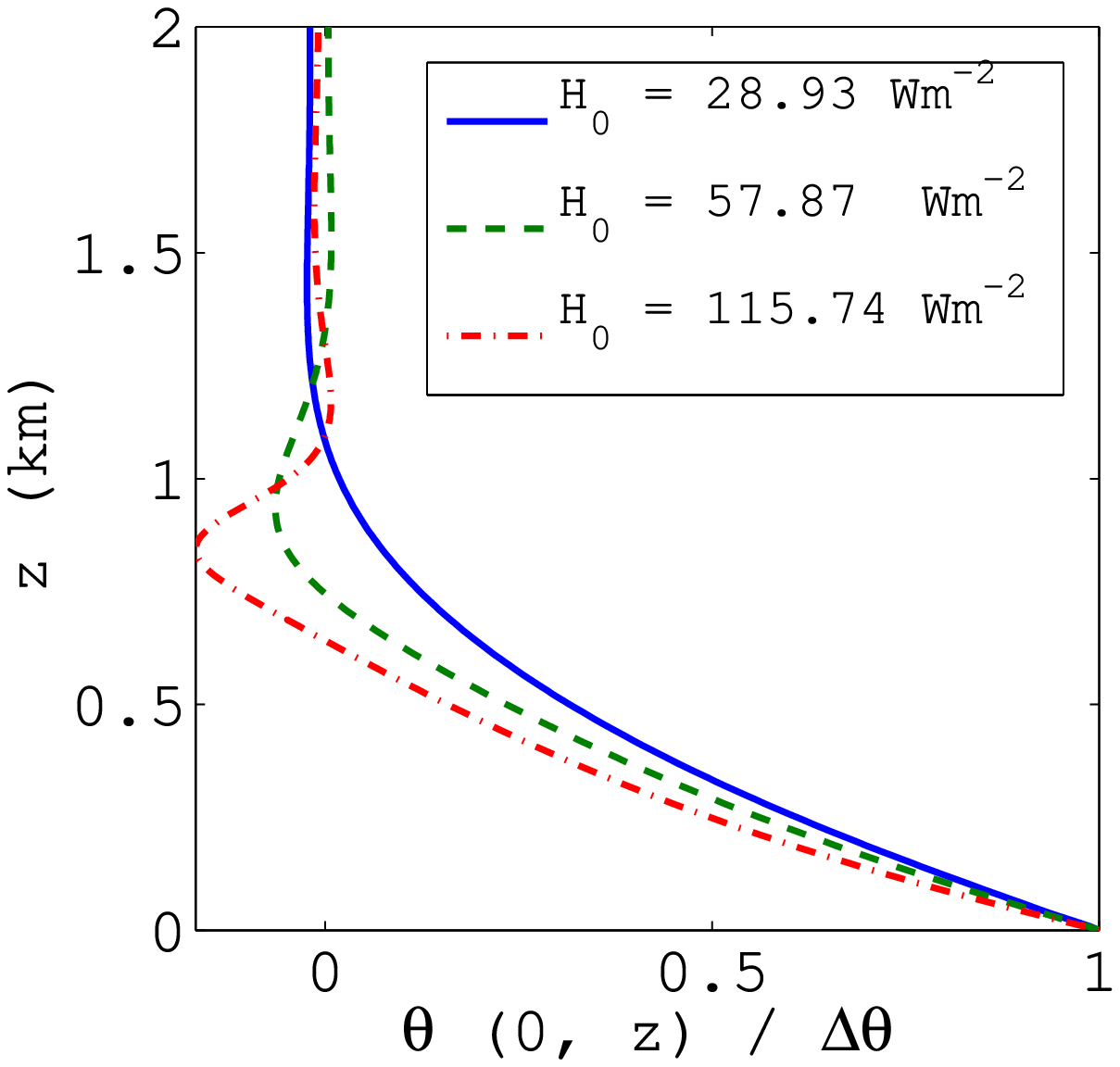}
   \caption{}
   \label{fig:comparison_theta_dubois}
 \end{subfigure}\quad %
 ~ 
 \begin{subfigure}[b]{0.31\textwidth}
   \includegraphics[trim=1cm 0cm 1.5cm 0cm, width=\textwidth]{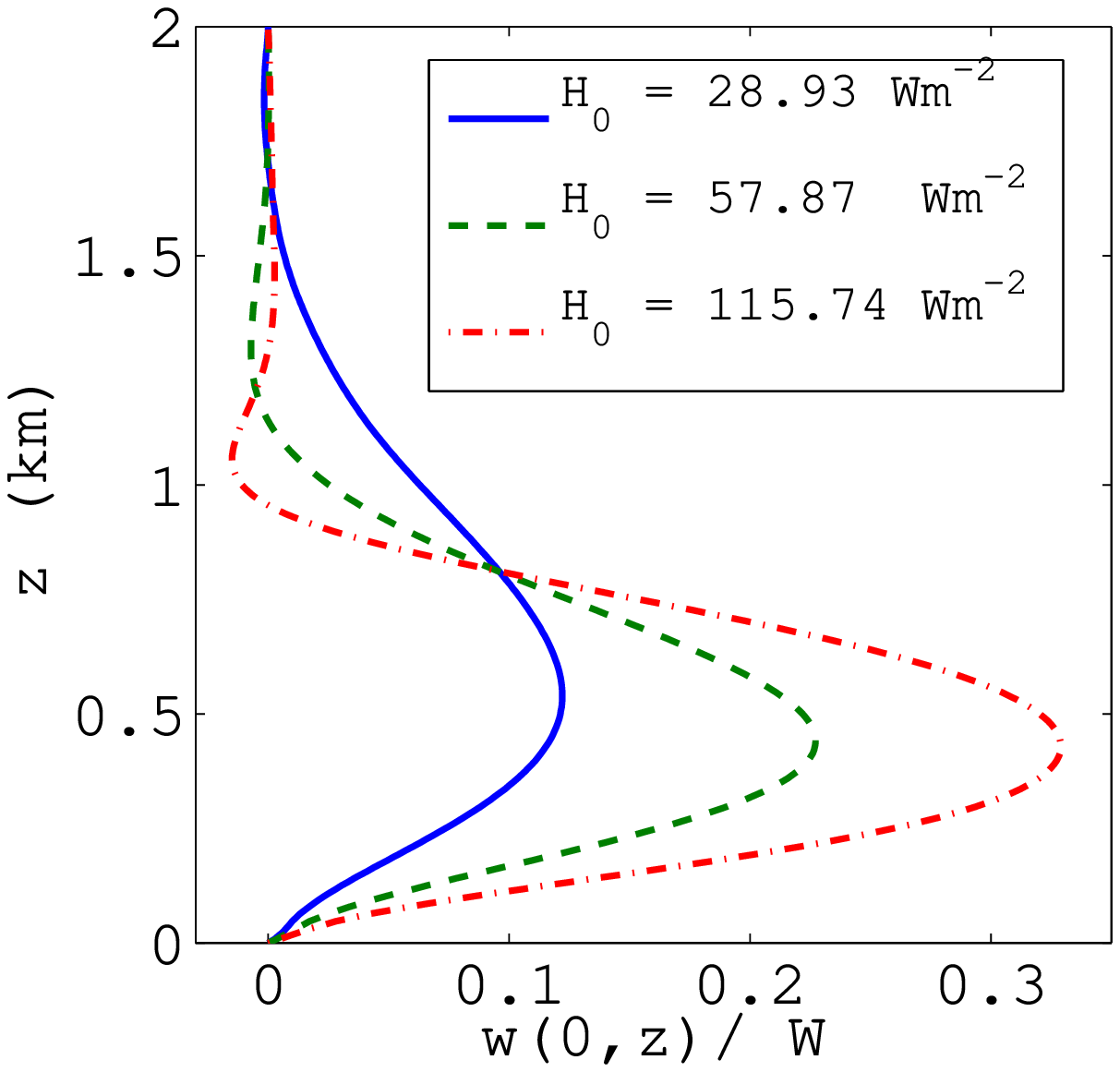}
   \caption{}
   \label{fig:comparison_theta_present}
 \end{subfigure}
~
 \begin{subfigure}[b]{0.32\textwidth}
   \includegraphics[trim=1cm 0cm 1cm 0cm, width=\textwidth]{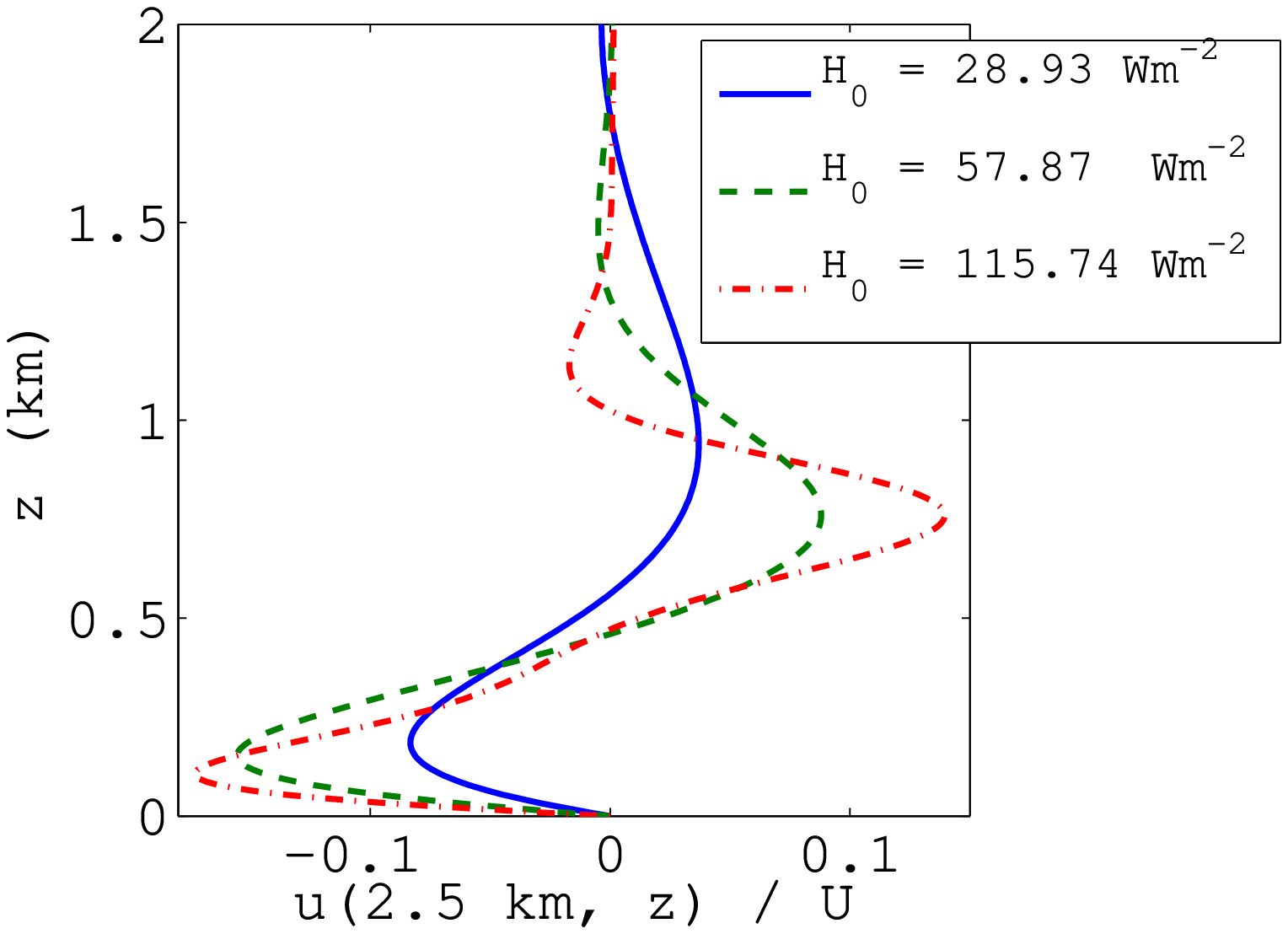}
   \caption{}
   \label{fig:du25}
 \end{subfigure}
 
 \caption{Profiles of (a) temperature variation at $x = 0$ km, (b) vertical velocity at $x=0$ km and (c) horizontal velocity at $x = 2.5$ km  at 6.5 h. \label{fig:comparison_dt}}
\end{figure*}

\section{Numerical Simulation and Validation \label{sec:bc}}
The design of this UHI simulation is analogous to the large eddy simulation presented by \cite{Zhang2014}. A stably stratified atmosphere is initialized in which the average potential temperature $\bar \theta$ increases linearly with height i.e. $\bar \theta = \theta_0 + \beta z$, where $\theta_0$ is a reference potential temperature, and $\beta$ is a constant vertical gradient of potential temperature. The heating of the surface owing to the sun in the morning generates a heat flux toward the air. The nocturnal stable boundary layer tends to diminish due to this surface heat flux as well as the above airflow. Finally, a mixed layer will be produced due to turbulence by means of high intensity heat flux and mixed air flow which is shown schematically in Fig. \ref{fig:uhisv}.  The urban heat island circulation is a horizontally dominated converging and diverging circulation. Plumes in Fig. \ref{fig:uhisv} indicate the direction of converging and diverging flow due to UHI effect.

The domain of the present simulation is considered $[-\frac{L_x}{2}, ~\frac{L_x}{2}] \times [0, ~L_z]$ and the domain of the heated region is considered $[-l, ~l]$ in which surface temperature is $\Delta \theta$ higher than the surrounding rural region, which is shown in Figure \ref{fig:uhisv}. Boundary conditions at the surface, $z = 0$, are: $u = w = 0$. The surface potential temperature perturbation $\theta_s$~($\hbox{\tt K}$) over the ground area ($z = 0$) is defined by 
 \[\theta_s(x) = a\left(\tanh\left(\frac{bx+1}{\xi}\right)-\tanh\left(\frac{bx-1}{\xi}\right)\right)\]
where $a = 0.5,~b = 1,~\rm{and} ~\xi = 0.01$ are constants. 

We have chosen open boundary conditions except the surface, where at $x = \pm L_x/2$, $\frac{\partial u}{\partial x}=0$, $\frac{\partial w}{\partial x} = 0$ and $\frac{\partial \theta}{\partial x}=0$, and at $z=L_z$, $\frac{\partial w}{\partial z}=0$, $\frac{\partial u}{\partial z} = 0$ and $\frac{\partial \theta}{\partial z}=0$. The upper boundary conditions allow transmission of internal waves, which is generated in a stable layer of the atmosphere. In \cite{Dubois2009}, a local damping technique is applied at the vertical boundaries; however, the present model can manipulate the simulations without any artificially imposed layer along the boundaries. 

The simulation domain is $L_x = 100$ km long in the horizontal direction, $L_z = 2$ km wide in the vertical direction and the heated region either $10$ km or $20$ km long. In numerical simulations, six different surface heat fluxes are considered in the heated region to examine the flow regime of UHI. The surface heat fluxes in those six situations are $H_0 = 28.93, 57.87, 115.74$, $231.48, 462.96$ and  $925.92$ W m$^{-2}$. The results between 6 h to 8 h of these six cases are analyzed in the following sections.
\section{Numerical Method}
A weighted residual collocation method is employed to represent the resolved flow with the basis $\{\varphi_k(x)\}_{k=0}^N\}$, where $\varphi_k(x)$,s are Deslauriers-Dubuc (DD) interpolating scaling functions. In \cite{Alam2014}, the authors developed a numerical technique to discretize partial derivatives using the filters that are used to construct $\varphi_k(x)$. In this article, we have used such a basis with a vanishing moment of 4 which means that the support of each scaling function is 7 grid intervals. The time integration is obtained by a Krylov method, where the Jacobian free Newton-Krylov (JFNK) approach has been adopted. Note that the classical JFNK solver requires a preconditioner. In the present work, DD discretization serves for an indirect preconditioning and we have not adopted any explicit preconditioning. On a $1025 \times 513$ mesh, the present solver takes about 1 to 5 Newton iterations where the first 1 to 3 iterations require about 5 GMRES iterations. One important advantage of the method is that the scheme has no mechanism of artificial damping (see Fig. 8 in \cite{Alam2014}). The time steps can be chosen independently of the CFL restriction. The classical Smagorinsky model is adopted to approximate the subgridscale process.  

\section{Results and discussion}
\begin{table*}[t]
\begin{center}
\begin{tabular}{crrrrrr}
\hline\hline
$H_0$ & \multicolumn{2}{ c }{$28.93$ W m$^{-2}$}& \multicolumn{2}{ c }{ $57.87$ W m$^{-2}$} &\multicolumn{2}{ c }{ $115.74$ W m$^{-2}$}\\
\hline
&Present&D \& T&Present&D \& T&Present&D \& T\\
\hline
$\theta_{\min}$ & -0.023537  & -0.024823& -0.064457 &-0.071289 &  -0.167264  & -0.166316\\ 
$\theta_{\max}$ & $1.0$ &1.0 & $ 1.0$ &1.0 & 1.0 &1.0\\
$u_{\max}$ & 0.118872 &0.118887 &  0.176103 &0.174844 & 0.179622 &0.179054\\
$w_{\min}$ &  -0.030134 & -0.030470& -0.037337  & -0.039291&   -0.085519 &-0.079265\\
$w_{\max}$ &  0.122229 & 0.125594& 0.227591 & 0.228250&   0.329467 & 0.322483\\
$\omega_{\max}$ & 1.957423 & 2.06900 & 3.659917 & 3.951325 & 5.345340 & 5.921375 \\
$Nu$ & 0.147744 & 0.148605 & 0.326211 & 0.295132 & 0.689294 & 0.643594 \\
\hline
\end{tabular}
\end{center}
\caption {Comparison of extreme values with \cite{Dubois2009} (D \& T) method} 
\label{tab:comparison_dubois}
\end{table*}

\subsection{Dependence of Flow Characteristics at Small Surface Heat Fluxes}\label{comparison_dubois}
To observe the flow regime at relatively small values of $H_0$ and to compare the solution with a reference model, three values of $H_0$ are considered such as $28.93, 57.87$ and $115.74$ W m$^{-2}$. These heat fluxes correspond to Rayleigh number $\mathrm{Ra} = 10^3, 10^4$ and $10^5$, respectively. The results are analyzed in comparison with that were presented by Dubois and Touzani \cite{Dubois2009} for the same values of $\mathrm{Ra}$. As it was observed in \cite{Dubois2009}, the flow reaches a fully developed and statistically stationary state within $2$~h from the initialization. Thus for $H_0 \le ~115.74$ W m$^{-2}$, the present simulations represents fundamental characteristics of UHI.   

For a more quantitative understanding, the vertical profiles of numerical solution from the present model can be compared with that from \cite{Dubois2009}. In Figure \ref{fig:comparison_dt}, $\theta (0\,\textrm{km}, z, 6.5\,\textrm{h})$, $w (0\,\textrm{km}, z, 6.5\,\textrm{h})$, and $u (2.5\,\textrm{km}, z, 6.5\,\textrm{h})$ are compared for three values of $H_0$. To help comparison with \cite{Dubois2009}, the solution in Fig. \ref{fig:comparison_dt} are normalized. When these plots are compared with that presented in Fig. 9 and 11 of \cite{Dubois2009},  one observes an excellent agreement although numerical methods are different. For each figure, a solid line, dashed line and dashed-dotted line represent the profiles for $H_0 = 28.93, ~57.87$ and $115.74$ W m$^{-2}$, respectively.  The vertical propagation of perturbations is controlled by the vertical stratification. In Figure \ref{fig:comparison_dt}b, The vertical velocity is maximum at $z \sim 0.5$~km, and it becomes larger for larger surface heat flux. In Figure \ref{fig:comparison_dt}c, it is like pollution will be transported toward the city at lower level and away from the city at upper level, as demonstrate schematically in Fig. \ref{fig:uhisv}. The converging and diverging flows are the maximum at $z \sim 0.2$~km and $z \sim 0.8$~km, respectively. In the horizontal direction, the vertical velocity vanishes rapidly outside the heated region. The temperature perturbation, vertical velocity and horizontal velocity decay rapidly above the mixed layer, as shown in Figure \ref{fig:comparison_dt}. In Figure \ref{fig:comparison_dt}a and \ref{fig:comparison_dt}b, it is clear that the mixed layer height appears between $z = 1.2~ \sim ~1.5$ km, and this height is reduced when surface heat flux becomes larger.

For the purpose of comparison with \cite{Dubois2009}, the nondimensionalized extreme values of the horizontal velocity $u$, the vertical velocity $w$, the temperature perturbation $\theta$, the vorticity $\omega = \partial w/\partial x - \partial u/\partial z$ and the Nusselt number $Nu$ are given in Table \ref{tab:comparison_dubois}.  The magnitude of the extreme values of $u$, $w$, $\omega$ and $Nu$ are increased as a function of surface heat flux. In Table \ref{tab:comparison_dubois}, magnitude of the vertical velocity increases significantly as a linear relationship, and it is clear that the surface heat flux is a limiting factor to accelerate vertical velocity ($w$).

\begin{figure*}
  \centering             
 \begin{subfigure}[b]{0.45\textwidth}
   \includegraphics[width=\textwidth]{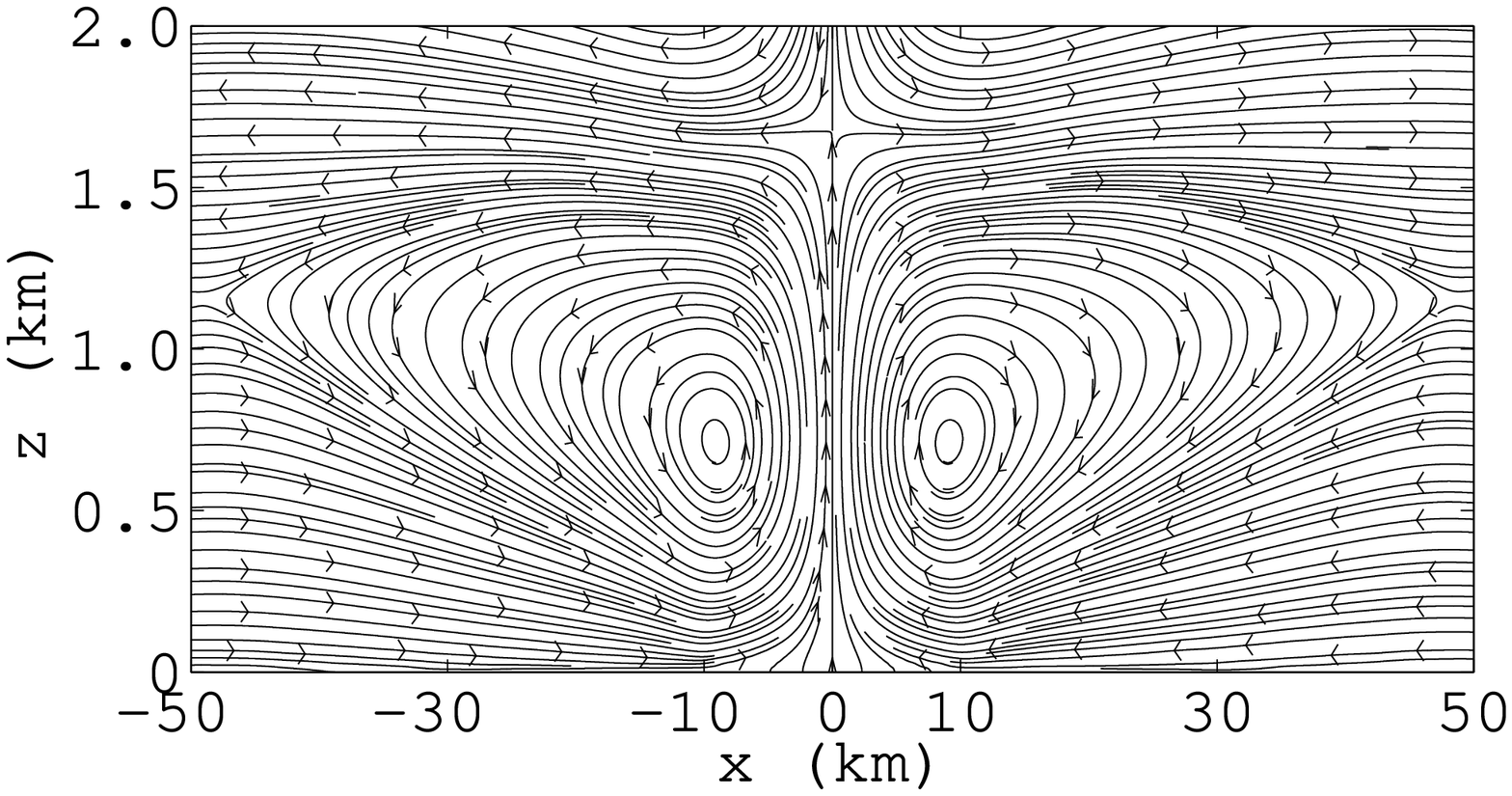}
   \caption{}
   \label{fig:stream_ra3}
 \end{subfigure}\quad %
 ~ 
 \begin{subfigure}[b]{0.45\textwidth}
   \includegraphics[width=\textwidth]{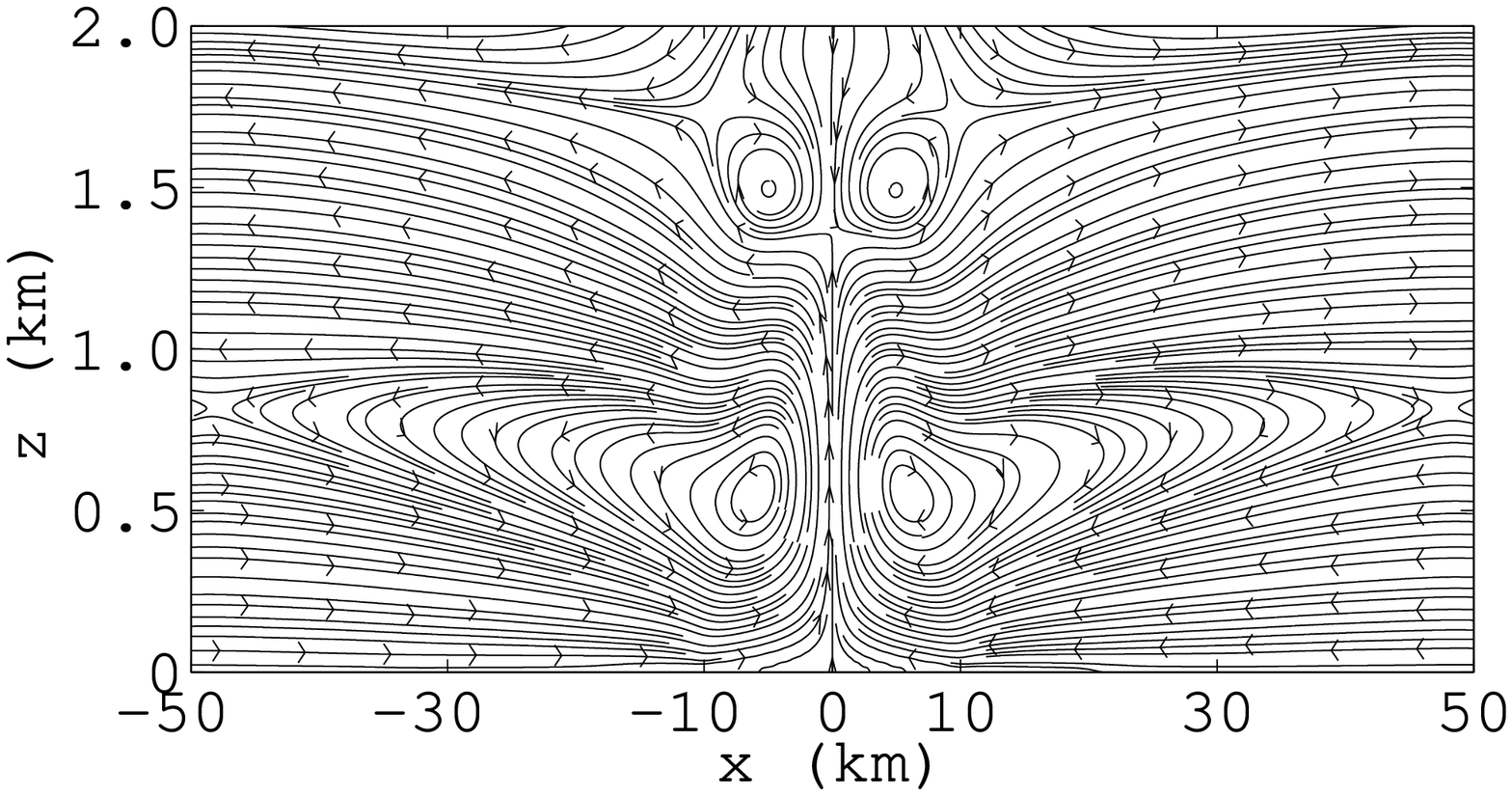}
   \caption{}
   \label{fig:stream_ra4}
 \end{subfigure}
~
 \begin{subfigure}[b]{0.45\textwidth}
   \includegraphics[width=\textwidth]{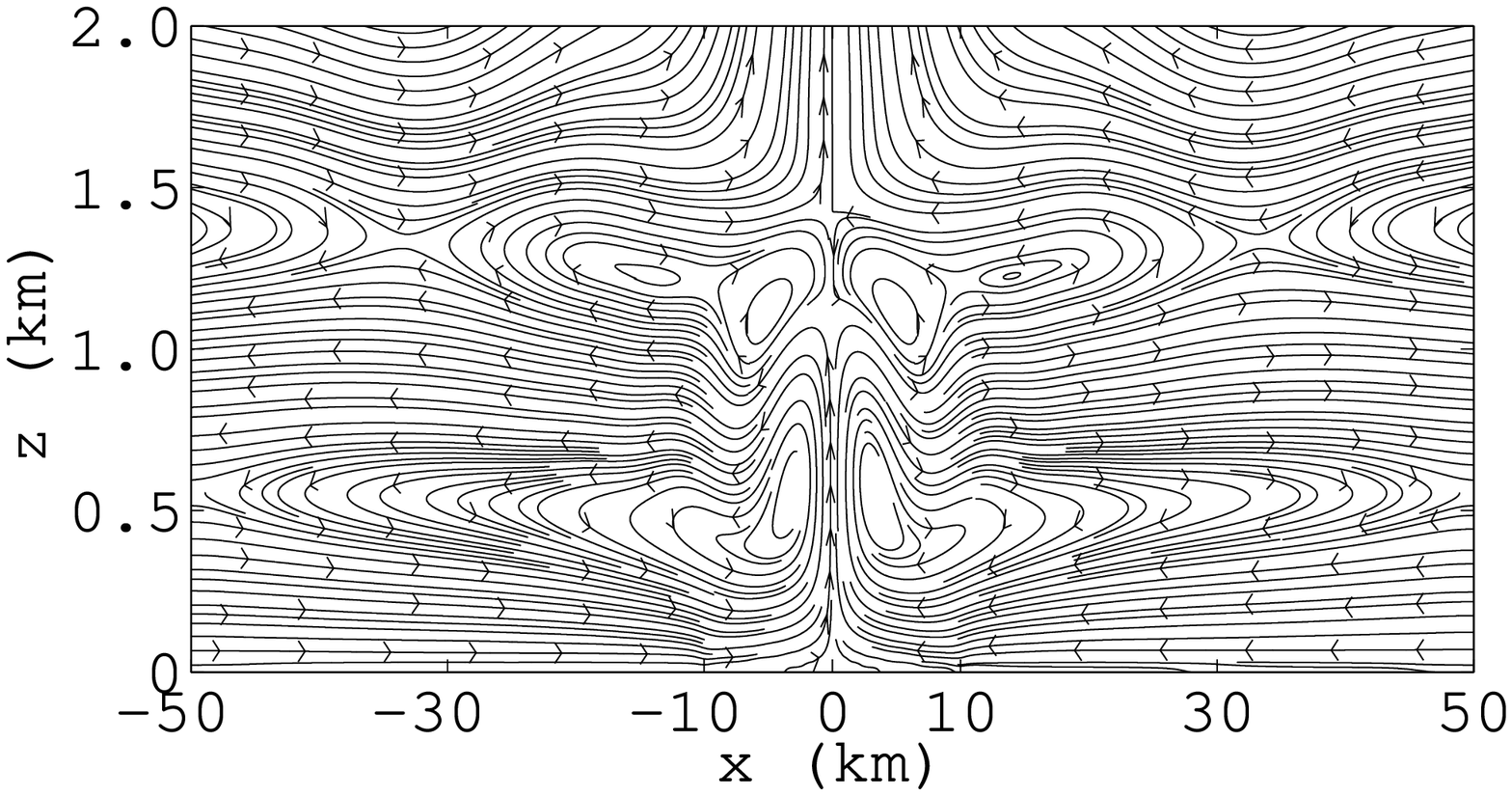}
   \caption{}
   \label{fig:stream_ra5}
 \end{subfigure}
 
 \caption{Streamlines for the UHI circulations for (a) $H_0 = 28.93$ W m$^{-2}$, (b) $H_0 = 57.87$ W m$^{-2}$ and (c) $H_0 = 115.74$ W m$^{-2}$ at $t = 6.5$ h. \label{fig:comparison_stream}}
\end{figure*}

Heat island circulation exibits two types of flow regimes \cite{Kimura1975, Nino2005}. When the differential heating is weak, the centre of the circulation is located at the edges of the heat island and the up-draft prevails all over the heat island. On the other hand, when the differential heating is strong, a strong narrow up-draft is concentrated above the centre of the island. The flow regime and the location of the centre of the circulation depends on the initial surface heat flux. When the surface heat flux increases, the center of the circulation moves toward the center of the island. For heat flux, $H_0 = 28.93,~57.87~\textrm{and}~115.74$ W m$^{-2}$ the stream lines of UHI circulations are shown in Figure \ref{fig:stream_ra3}, \ref{fig:stream_ra4} and \ref{fig:stream_ra5}, respectively.  The center of the circulation of Fig.~\ref{fig:stream_ra5} is close to the center of the island, and a strong updraft is observed. However, in Fig. \ref{fig:stream_ra3} the center of the circulation is far away from the center of island and the up-draft prevails due to weak heating. The overall pattern of the circulation is in agreement with what was presented by Kimura \cite{Kimura1975} and Nino and Mori \cite{Nino2005}.


\begin{figure*}[t]
\noindent\includegraphics[width=\textwidth]{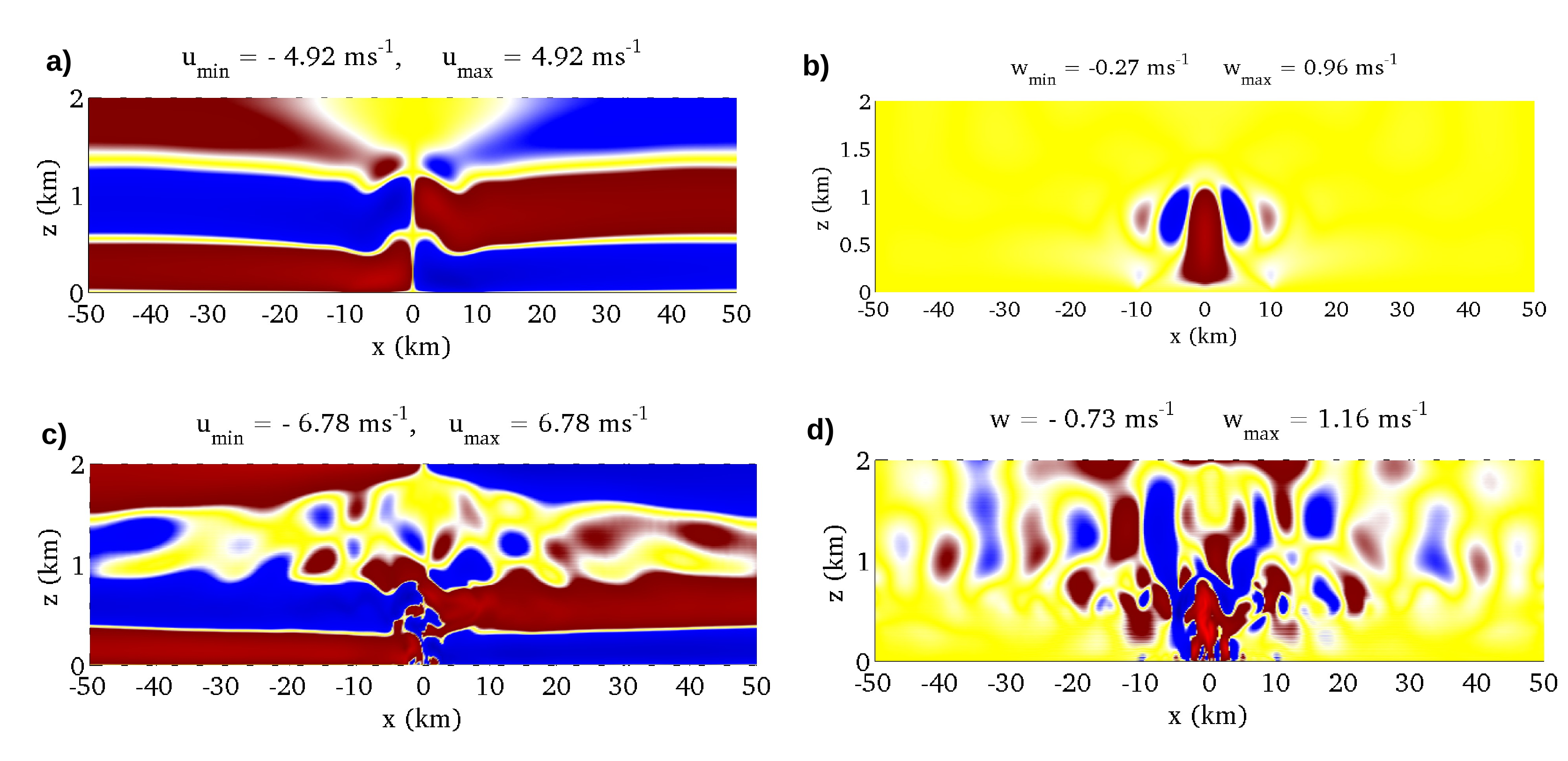}
 \caption{Horizontal velocity (left) and vertical velocity (right) for  (a) (b) $H_0 = 115.74$ W m$^{-2}$ and (c) (d) $H_0 = 462.96$ W m$^{-2}$ at $t=6.5$ h. Red, blue and yellow colors represent positive, negative and zero, respectively.}\label{fig:uvRa678}
\end{figure*}
\begin{figure*}[t]
\noindent\includegraphics[width=\textwidth]{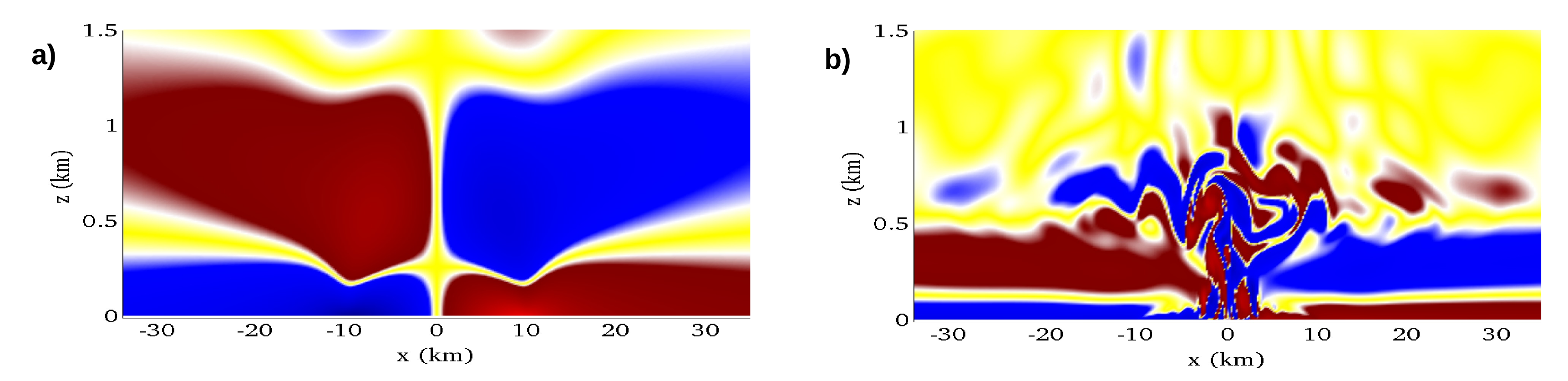}
 \caption{The formation of coherent structures at $t = 6.5$ h for (a) $H_0 = 28.93$ W m$^{-2}$ and (b) $H_0 = 462.96$~W~m$^{-2}$. Blue and red colors represent the clockwise rotation and the counter-clockwise rotation. }\label{fig:vorticity_Ra3_8_Ri1} 
\end{figure*}

\subsection{Dependence of Flow Patterns for Large Surface Heat Flux}
A strong differential heating creates a strong narrow up-draft near the centre of the island. This narrow up-draft causes instabilities for large surface heat flux, and finally flow regime becomes turbulent. Figure \ref{fig:uvRa678} compares horizontal velocity and vertical velocity for surface heat fluxes, $H_0 = 115.74,~462.96$ W m$^{-2}$ at $t = 6.5$~h. Clearly, above the surface layer, the flow is towards the center of the UHI (the converging wind); conversely, further above the surface layer the flow is outward (the diverging wind). For the cases $H_0 \le 231.48$ W m$^{-2}$, the fully developed flow pattern is symmetric about the center of the heat island ({\em e.g.} Fig. \ref{fig:uvRa678}a). When surface heat flux $H_0 > 231.48$ W m$^{-2}$, symmetry of the horizontal velocity is broken and turbulent flow is visible in the heat island and its vicinity ({\em e.g.} Fig. \ref{fig:uvRa678}c). For $28.93 \le H_0 \le 231.48$ W m$^{-2}$, vertical velocity concentrates only in the urban area ({\em e.g.} Fig. \ref{fig:uvRa678}b). On the otherhand, for larger surface heat flux, $H_0 \ge 462.96$ W m$^{-2}$ vertical velocity appears throughout the urban and rural area ({\em e.g.} Fig. \ref{fig:uvRa678}d). 

The large UHI heat intensity causes stronger convergence and divergence wind and both the horizontal and vertical wind speeds depend on the UHI intensity. The outward wind speed is much higher than the inward wind speed in the convective boundary layer and the vertically upward wind speed is higher than the downward wind speed. The maximum outward wind speeds for $H_0 = 231.48$, $462.96$ and $925.92$ W m$^{-2}$ are $5.19$, $6.78$ and $8.21$ m s$^{-1}$, respectively, whereas the maximum inward wind speeds are $3.37$, $4.43$ and $5.36$ m s$^{-1}$ respectively. Similarly, the maximum upward wind speeds for $H_0 = 231.48$, $462.96$ and $925.92$ W m$^{-2}$ are $1.08$, $1.16$ and $1.31$ m s$^{-1}$, respectively, whereas the maximum downward wind speeds are $0.49$, $0.73$ and $0.85$ m s$^{-1}$, respectively. 

\subsection{Coherent Turbulent Structure}
Coherent vortices are an efficient and accurate basis to analyze turbulent flow. In Figure \ref{fig:vorticity_Ra3_8_Ri1}, vorticity is depicted to analyze the coherent structure of turbulent flow. For a large surface heat flux, the initial turbulent eddies are created in the urban-rural interface, and they turn-over in this interface. After time being convective velocity grow up; as a result, eddies are created in the whole urban area and the turbulent mixing spread the rural areas. These continuously changing vortices that move from urban to rural areas represent a turbulent state which is illustrated by the Figure \ref{fig:vorticity_Ra3_8_Ri1}b. Vorticity structures show that the flow is not turbulent for $28.93$~W~m$^{-2} \le H_0 \le 231.48$~W~m$^{-2}$ ({\em e.g.} Fig. \ref{fig:vorticity_Ra3_8_Ri1}a), and is turbulent for $H_0 = 462.96$ and $925.92$~W~m$^{-2}$ ({\em e.g.} Fig. \ref{fig:vorticity_Ra3_8_Ri1}b). These changes are responsible for the increased intermittency of high-Rayleigh number turbulence.  

\subsection{Potential Temperature and Mixed-layer Height}
\begin{figure}[t]
\noindent\includegraphics[width=7cm,height=10cm]{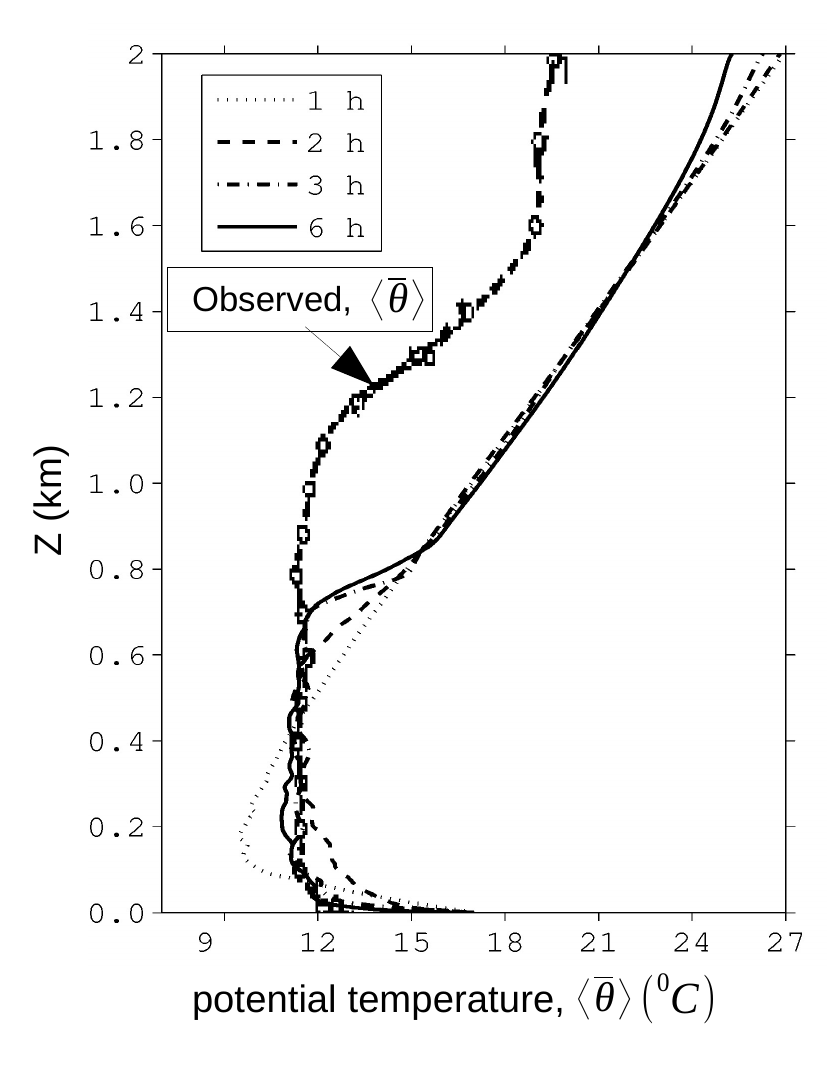}\\
\caption{A comparison of $\langle\bar \theta\rangle$ with the field measurment of the CBL profile (e.g. \cite{moeng1984large}). \label{fig:uhi}}
\end{figure}

The vertical profile of the potential temperature is an important characteristic to understand atmospheric layers. In day time, the surface layer is unstable, the mixed layer is neutral, and the entrainment zone and the atmosphere above are stable ({\em e.g.} \cite{deardorff1969laboratory, moeng1984large}).  We have analyzed the daytime boundary layer pattern for the present UHI simulation. Since the present study is not site-specific, we have compared $\langle\bar\theta\rangle$ obtained from the present simulation with that observed by the Wangara experiment ({\em e.g.} \cite{moeng1984large}). The result presented in Figure \ref{fig:uhi} for the surface heat flux $H_0 = 925.92$ W m$^{-2}$ shows the development of a well-mixed layer by 2~h, which is neutral. The mixed layer height in the turbulent region at $\sim 800$~m is capped with the inversion layer as shown in Fig. \ref{fig:uhi}. At the bottom of the boundary layer, the unstable surface layer appears at $\sim 50$~m. A comparison of the vertical profile of potential temperature with the field measurement obtained from the Wangara day 33 experiments ( {\em e.g.} \cite{moeng1984large}) shows strong agreement between the two data set in the mixed layer. The mixed layer height due to the surface heating may rise for an additional shear turbulence owing to a synoptic scale wind.

\subsection{Internal Wave Generation}
\begin{figure}[t]
  \noindent\includegraphics[width=18pc,angle=0]{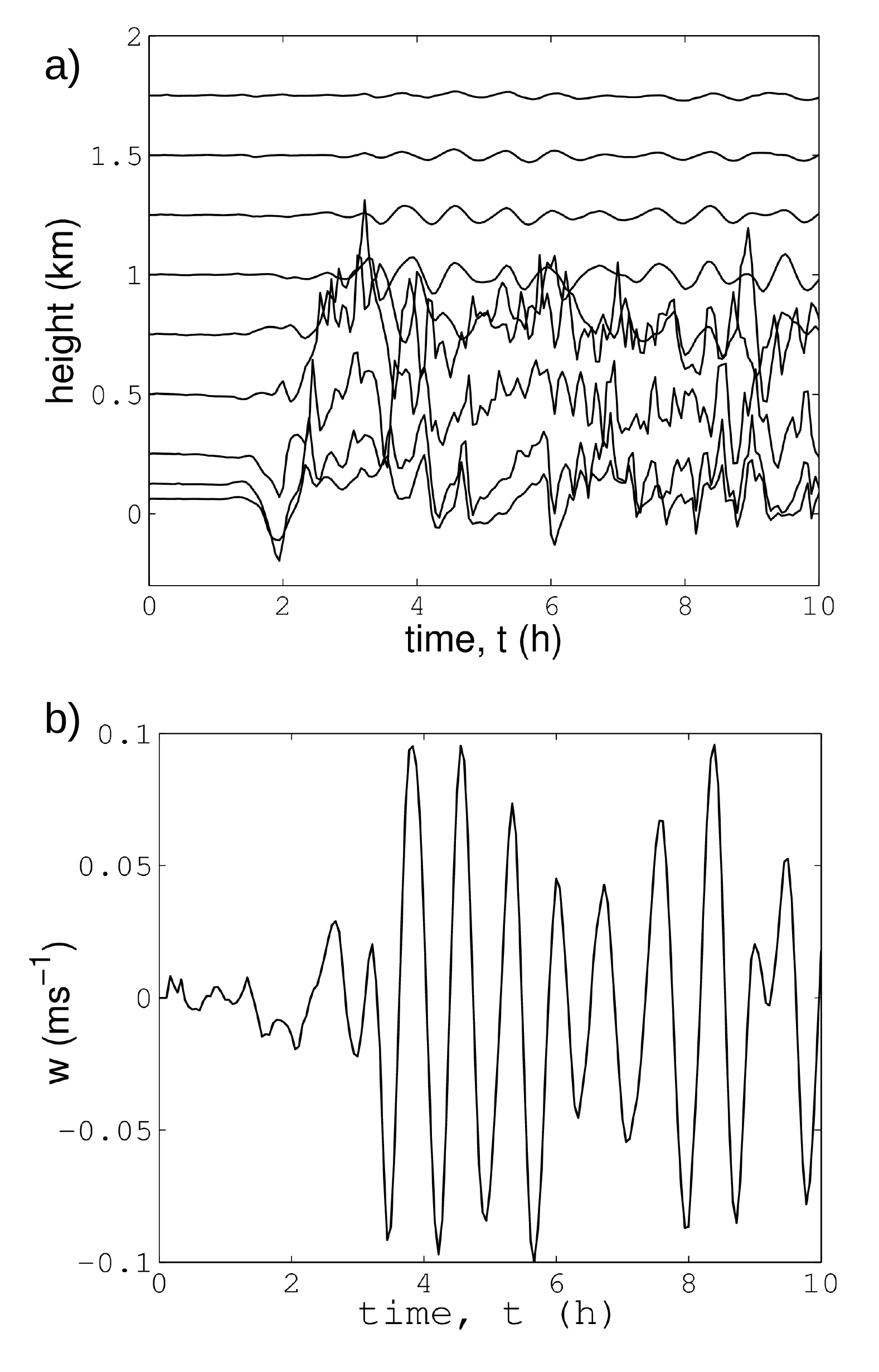}
  \caption{ Generation of internal waves for $H_0 = 462.96$~W~m$^{-2}$. (a) Evolution of vertical velocity at different heights along the center of the heat island. (b) internal wave at height $z = 1.5$~km.  \label{fig:ch5_wave}}
\end{figure}

The turbulent convection in the ABL generates internal waves in the stable layer that overlays the mixed layer during the daytime. Surface heating in daytime creates thermals that penetrate upward and bounce back and as a result, vertical velocity fluctuates inside the mixed layer, which creates internal waves in the stable atmosphere ({\em e.g.} \cite{Lane2008}). In Figure \ref{fig:ch5_wave}a, the evolution of vertical velocity at different heights of the center of the heat island is presented for $H_0 = 462.96$~W~m$^{-2}$. To observe the internal wave in the free atmosphere, an internal wave at height $z = 1.5$~km is shown in Figure \ref{fig:ch5_wave}b for $H_0 = 462.96$~W~m$^{-2}$. These internal waves can transfer a portion of energy from the mixed layer of the ABL to the upper layer of the stable atmosphere. Further study of internal waves due to heterogeneity of surface heat flux is in progress.\\

\section{Conclusions}
Flow regimes of mesoscale circulations have been investigated for a differentially heated large horizontal area. The circulations are principally governed by the difference of surface heat fluxes between the urban and rural areas. A range of surface heat fluxes, $H_0 = 28.93 - 925.92$~W~m$^{-2}$ have been considered to investigate flow regimes. For lower surface heat fluxes, flow regimes have been compared with the reference numerical and experimental results. Analyzing the flow regimes and coherent vortices, it is concluded that the flow exhibits steady and symmetric multi-cell pattern centered above the UHI region for $28.93$~W~m$^{-2}$ $\le H_0 \le 115.74$~W~m$^{-2}$. The stability of steady states is lost for large surface heat flux. Flow regime is symmetric but not steady for $H_0 = 231.48$~W~m$^{-2}$ and turbulent for $H_0 = 462.96$~W~m$^{-2}$ and $925.92$~W~m$^{-2}$. For the turbulent case, the vertical profiles of potential temperature have been analyzed in which mixed layer heights appear at the place where the flow is turbulent. Due to the penetration of turbulence in the stable layer, internal  waves appear which transfer energy from the ABL to the free atmosphere. 

Though the real atmosphere is a moist atmosphere with a synoptic scale wind component and the governing equations are simplified with several assumptions, the present results give a good framework for understanding the dynamics of heat island circulation for a wide range of surface heat flux. Dependence of simulations for moisture and synoptic scale wind component will be addressed in future works.


\bibliography{Thesis}

\end{document}